\documentclass[aps,twocolumn]{revtex4}
\usepackage[usenames,dvipsnames]{color}

\usepackage{graphicx}
\usepackage{amssymb}
\usepackage{subfigure}
\usepackage{stackrel}
\usepackage{enumitem}
\usepackage{amsmath}
\usepackage{amsfonts}
\usepackage{bm}
\usepackage{color}
\usepackage{verbatim} 
\usepackage{tcolorbox}
\usepackage[normalem]{ulem}
\usepackage{hyperref}
\usepackage{soul}
\usepackage{mathtools}

\newlist{subquestion}{enumerate}{1}
\setlist[subquestion,1]{label=(\alph*)}

\newcommand{\ket}[1]{\ensuremath{\left|{#1}\right\rangle}}

\definecolor{rosita}{rgb}{0.97, 0.56, 0.65}

\newcommand{\sd}[2]{\ensuremath{|\psi^{sl}_{{#1}{#2}}\rangle}}

\newcommand{\beq}{\begin{equation}}
\newcommand{\eeq}{\end{equation}}
\newcommand{\bse}{\begin{subequations}}
\newcommand{\ese}{\end{subequations}}\newcommand{\bea}{\begin{eqnarray}}
\newcommand{\eea}{\end{eqnarray}}
\newcommand{\bit}{\begin{itemize}}
\newcommand{\eit}{\end{itemize}}
\newcommand{\bpmatrix}{\begin{pmatrix}}
\newcommand{\epmatrix}{\end{pmatrix}}

\newcommand{\be}{\begin{equation}}
\newcommand{\ee}{\end{equation}}
\newcommand{\ben}{\begin{eqnarray}}
\newcommand{\een}{\end{eqnarray}}

\begin{document}

\title{Speed of evolution in entangled fermionic systems}
\author{Sahory Canseco J.}
\author{Andrea Vald\'{e}s-Hern\'{a}ndez}
\email[]{andreavh@fisica.unam.mx}
\thanks{corresponding author}
\affiliation{Instituto de F\'{\i}sica, Universidad Nacional Aut\'{o}noma de M\'{e}xico, Apartado Postal 20-364, Ciudad de M\'{e}xico, Mexico.}

\begin{abstract}
We consider the simplest identical-fermion system that exhibits the phenomenon of entanglement (beyond exchange correlations) to analyze its speed of evolution towards an orthogonal state, and revisit the relation between this latter and the amount of fermionic entanglement. A characterization of the quantum speed limit and the orthogonality times is performed, throwing light into the general structure of the faster and the slower states. Such characterization holds not only for fermionic composites, but apply more generally to a wide family of 6-dimensional states, irrespective of the specific nature of the system. Further, it is shown that the connection between speed of evolution and entanglement in the fermionic system, though more subtle than in composites of distinguishable parties, may indeed manifest for certain classes of states.

 
\end{abstract} 


\maketitle

\section{Introduction\label{sec:intro}}

The simple question on whether a state of a given quantum system can evolve towards an orthogonal (distinguishable) one, has led to remarkably fruitful lines of research that deepen into the dynamics of quantum systems and into the ultimate limits of quantum devices. The interpretation of the energy-time uncertainty relation advanced by Mandelstamm and Tamm \cite{Mandelstam1991}, followed by the work of Margolus and Levitin \cite{MargolusLevitinPhysD1998}, led to the actual notion of \emph{quantum speed limit} ($\tau_{\textrm{qsl}}$) as the lower bound for the \emph{orthogonality time} $\tau$ required for an initial state to evolve towards an orthogonal one. Quantum information, quantum control, quantum metrology, quantum thermodynamics, among other areas of current research, rely on such bound to establish the fundamental limits that restraint quantum processes in general and the profit we can get by exploiting them (see  \cite{DeffnerJPA2017,FreyQInfProcc2016} and references therein).

Despite its intrinsic fundamental and practical significance, when considering time-independent Hamiltonian evolutions, the quantum speed limit can be attained only by equally-weighted coherent superpositions of two (non-degenerate) energy eigenstates, that is, by pure states of effective 2-level systems, or qubits, with a specific energy distribution \cite{LevitinPRL2009,BasonNat2012}. 
In general, pure states of a $n$-level autonomous system do not even reach an orthogonal state ---they do so provided certain conditions are met both by the Hamiltonian and the probability distribution of the energy eigenstates \cite{ValdesJPA2020}---, and when they do the orthogonality time may approximate, though not reach, the quantum speed limit \cite{NessScienceAdv2021}. The characterization of the initial states that transform into an orthogonal one in a finite time for a given (time-independent) Hamiltonian, and the identification of those that do it faster, is a problem that remains open for arbitrary $n$ (a comprehensive characterization of the states that reach orthogonality has been addressed in \cite{SevillaQRep2021} for qutrit systems, $n=3$), and acquires relevance in the study of the dynamics of multi-level systems, as well as in the preparation of states for specific information tasks.     

A particularly interesting vein in the problem of states that evolve into a distinguishable one, was revealed by Giovanetti et al. \cite{GiovannettiEPL2003,GiovannettiPRA2003} who showed that when bipartite quantum systems are considered, entanglement can speed up the evolution towards orthogonality. The relation between entanglement and the speed of evolution has been analyzed in relation with the brachistochrone problem \cite{BorrasEPL2007,BorrasPRA2008,ZhaoPRA2009,CarliniJPA2017}, in $N$-qubit \cite{ZanderJPA2007,LiuPRA2015,FrowisPRA2012,ZhangSciRep2016}, 2-qubit \cite{BorrasPRA2006,KupfermanPRA2008,BatlePRA2005,BatleErratum,ChauPRA2010} and 2-qutrit systems \cite{BehzadiAnnPhys2017}, and also in relation with entanglement measures \cite{RudnickiPRA2021}. A common feature of these works (except \cite{BorrasPRA2008,BatlePRA2005}) is that they focus only on systems composed of distinguishable constituents. By considering instead composites of indistinguishable parties, the (anti)symmetry of the states gives rise to a modification of the very notion and definition of entanglement between (identical) particles \cite{GhirardiJSP2002,EckertAnnPhys2002}, and the conclusions reached when studying distinguishable-particle states that evolve towards an orthogonal state may no longer hold. An analysis of the effects of entanglement on the quantum speed in bipartite systems of indistinguishable parties has been advanced for bosonic \cite{BatlePRA2005,BorrasPRA2008} and fermionic \cite{BatlePRA2005,BorrasPRA2008,OliveiraIJQ2008} systems. Interestingly, while in a system of non-interacting bosons a clear correlation between entanglement and evolution speed can be identified, in the analogous fermionic case such correlation is not observed \cite{BatlePRA2005}, yet it can emerge for a specific energy spectrum  \cite{OliveiraIJQ2008}. Further, the relation between entanglement and brachistochrone
evolutions is weaker in fermionic systems, in comparison with the identical-boson and the distinguishable-qubit case \cite{BorrasPRA2008}.

In order to enrich our comprehension of entangled fermionic systems that transit between orthogonal states, we consider a low-dimensional system of two non-interacting and identical fermions with a threefold purpose: i) to characterize the quantum speed limit of the system, ii) to identify the initial states that evolve into an orthogonal one in a finite time, and iii) to revisit the connection between entanglement and speed of evolution advanced in \cite{BatlePRA2005,OliveiraIJQ2008}, showing that such connection may indeed exist for certain classes of states. The present contribution throws light into the speed of evolution of the simplest fermionic system that exhibits the phenomenon of entanglement, and more generally on the speed of evolution of a wide family of pure states of a 6-level system, irrespective of its nature.  

The paper is organized as follows. In Sect. \ref{sec:Preliminaries} we go into the details of the specific system under consideration and introduce the measure of entanglement used throughout the work. Section \ref{ortogonal} lays out the \emph{orthogonality condition}, which imposes constraints on the initial states and on the time required to attain an orthogonal state. The energetic resources of the 6-dimensional two-fermion state are studied in Sect. \ref{secqsl}, and a characterization of the corresponding quantum speed limit is presented in terms of the relevant state's parameters. In Sect. \ref{examples} we focus on three paradigmatic types of initial states and explore their relation between $\tau$ and the amount of entanglement. In Sect. \ref{regions} a detailed analysis of the regions, in the relevant state's parameters space, consistent with the orthogonality condition is carried out, and the zones with greater and lower orthogonality times are identified. The connection between the evolution speed and entanglement is numerically explored in Sect. \ref{entanglement}, far beyond the examples studied in Sect. \ref{examples}. Finally, some final remarks are presented in Sect. \ref{conclusions}.

\section{The system under consideration} \label{sec:Preliminaries}

We will consider a pair of indistinguishable fermions, and denote with $\mathcal{H}_f$ the $d$-dimensional single-fermion Hilbert space, with an orthonormal basis $\{\ket{i}\}=\{\ket{1}, \ket{2},\dots, \ket{d}\}$. The Hilbert space of the composite system, denoted as $\mathcal{H}_-$, is the antisymmetric subspace of $\mathcal{H}_f\otimes\mathcal{H}_f$, with $\dim \mathcal{H}_-=d(d-1)/2$. A natural basis of $\mathcal H_-$ is given by the set $\{\sd{i}{j}\}$, where $\sd{i}{j}$ is the Slater determinant (SD for short)
\beq\label{slater}
\sd{i}{j}=\frac{1}{\sqrt{2}}(\ket{ij}-\ket{ji}),
\eeq 
with $i\neq j$, and $\ket{ij}=\ket{i}\otimes\ket{j}\in \mathcal{H}_f\otimes\mathcal{H}_f$.

In terms of the fermionic creation ($\hat f^{\dagger}_i$) and annihilation ($\hat f_i$) operators, which create and annihilate, respectively, a fermion in the $i$-th state and satisfy the anticommutation relations $\{\hat f_i,\hat f^{\dagger}_j\}=\delta_{ij}$, $\{\hat f_i,\hat f_j\}=0$, we can write
\beq
\sd{i}{j}=\hat f^{\dagger}_i\hat f^{\dagger}_j\ket{{\rm vac}}=\ket{n_1,n_2,\dots,n_{d}},
\eeq
where $\ket{n_1,n_2,\dots,n_{d}}$ stands for the Fock state satisfying $\hat f^{\dagger}_k\hat f_k\ket{n_1,n_2,\dots,n_{d}}=n_k\ket{n_1,n_2,\dots,n_{d}}$, with $n_k=1$ for $k=i,j$ and $n_k=0$ otherwise. The state $\ket{{\rm vac}}$ represents the vacuum (Fock) state having all $n_k=0$. 

An arbitrary pure state of the pair of fermions thus writes as  
\beq\label{psif}
\ket{\psi}=\sum^d_{i,j=1}w_{ij}\sd{i}{j}=\sum^d_{i,j=1}w_{ij}\hat f^{\dagger}_i\hat f^{\dagger}_j\ket{{\rm vac}},
\eeq
with $w_{ij}=-w_{ji}$. Since $\langle \psi^{sl}_{ij}|\psi^{sl}_{kl}\rangle=\delta_{ik}\delta_{jl}-\delta_{il}\delta_{jk}$, the normalization condition on $\ket{\psi}$ reads
\beq\label{norm}
\langle\psi|\psi\rangle=2\sum_{i,j}|w_{ij}|^2=4\sum_{i<j}|w_{ij}|^2=1.
\eeq

\paragraph*{Entangled fermionic states.-}
The fundamental vector $\sd{i}{j}$ represents a state with \emph{minimal} correlations, rooted at the exchange symmetry. Quantum correlations between the fermions other than those due only to their indistinguishability, are encoded in coherent superpositions of such minimally correlated states, that is, in superpositions of Slater determinants. These (extra) correlations correspond to what we identify, following \cite{GhirardiJSP2002}, as the entanglement between the identical fermions. In this approach, a pure state $\ket{\psi}$ of two indistinguishable fermions is regarded as non-entangled if and only if there exist a basis of $\mathcal {H}_-$ in which $\ket{\psi}$ is a minimally correlated state, hence expresses as a \emph{single} SD. If no such basis exists, the state is regarded as entangled
\cite{GhirardiJSP2002}. 

The two-fermion system of lowest dimensionality that may exhibit the phenomenon of entanglement (as defined above) corresponds to that in which each fermion has four accesible orthogonal states, so $d=4$ \cite{EckertAnnPhys2002}. In such case, a measure of the amount of entanglement in the generic pure state (\ref{psif}) is given by \cite{SchliemannPRB2001,EckertAnnPhys2002}
\begin{equation}\label{concurrencef}
 C_f(\ket{\psi})=8|w_{12}w_{34}-w_{13}w_{24}+w_{14}w_{23}|.
\end{equation} 
More generally, in terms of $\rho_f$, the reduced density matrix of a single fermion, the entanglement $C_f$ of a pure state of two $d$-dimensional fermions ($d\geq 4$) writes as \cite{MajteyPRA2016}
\beq
C_f(\ket{\psi})=\sqrt{\frac{2d}{d-2}\Big(\frac{1}{2}-\textrm{Tr}\rho^2_f\Big)}.
\eeq
We will refer to this quantity as the \textit{fermionic concurrence}, since it is an extension, to identical-fermion systems, of the usual concurrence quantifying the entanglement between two distinguishable parties in a pure state \cite{RungtaPRA2001}. The measure $C_f$ has been studied, for example, in connection with quantum walks of interacting fermions \cite{MelnikovSciRep2016}, fermionic quantum circuits \cite{GigenaPRA2017}, quantum dots \cite{SimonovicPRA2015}, and sudden death of entanglement in fermionic systems \cite{BussandriJPA2020}.

\paragraph*{Evolution of the system.-}
To characterize the simplest entangled fermionic states that evolve towards an orthogonal one, we will restrict our study to a pair of fermions with four accesible states each, and avoid any interaction between the fermions that could modify their entanglement as the evolution takes place. We thus consider an initial state given by (\ref{psif}) with $d=4$, and let it evolve unitarily with the non-interacting Hamiltonian
\beq
\hat H=\sum^{4}_{i=1}\epsilon_i\hat f^{\dagger}_i \hat f_i.
\eeq
The evolved state is therefore given by
\begin{eqnarray}\label{psift}
\ket{\psi(t)}&=&e^{-i\hat H t/\hbar}\ket{\psi(0)}\nonumber\\
&=&\sum^4_{i,j=1}w_{ij}e^{-i(\epsilon_i+\epsilon_j)t/\hbar}\sd{i}{j}.
\end{eqnarray}

Assuming further that the (single-fermion) energy levels $\epsilon_k$ are equally spaced, so that $\epsilon_k=k\epsilon$ with $k\in\{1,2,3,4\}$, the six accesible states of the composite system, and their corresponding energies $E$, are
\begin{subequations}\label{spectrum}
\begin{eqnarray}
\ket{\boldsymbol{1}}&\coloneqq& \sd{1}{2},\quad E_1=3\epsilon,\label{mine}\\
\ket{\boldsymbol{2}}&\coloneqq& \sd{1}{3},\quad E_2=4\epsilon,\\ 
\ket{\boldsymbol{3}}&\coloneqq& \sd{1}{4},\quad E_3=5\epsilon,\\ 
\ket{\boldsymbol{4}}&\coloneqq& \sd{2}{3},\quad E_4=5\epsilon,\\ 
\ket{\boldsymbol{5}}&\coloneqq& \sd{2}{4},\quad E_5=6\epsilon,\\ 
\ket{\boldsymbol{6}}&\coloneqq& \sd{3}{4},\quad E_6=7\epsilon.
\end{eqnarray}
\end{subequations}
This allows us to rewrite Eq. (\ref{psift}) as 
\beq\label{estadof}
\ket{\psi(t)}=\sum^6_{n=1}\sqrt{p_n}e^{i\theta_n}e^{-iE_nt/\hbar}\ket{\boldsymbol{n}}, 
\eeq
with $\{\theta_n\}$ arbitrary phases, and $p_n\in[0,1]$ the probability of the system being in the $n$-th state, satisfying $\sum_np_n=1$ and related to $\{w_{ij}\}$ according to
\begin{subequations}\label{ces}
\begin{eqnarray} 
2w_{12}&=&\sqrt{p_1}e^{i\theta_1}, \\ 
2w_{13}&=&\sqrt{p_2}e^{i\theta_2},\\ 
2w_{14}&=&\sqrt{p_3}e^{i\theta_3},\\
2w_{23}&=&\sqrt{p_4}e^{i\theta_4},\\
2w_{24}&=&\sqrt{p_5}e^{i\theta_5},\\
2w_{34}&=&\sqrt{p_6}e^{i\theta_6}.
\end{eqnarray}
\end{subequations}
In this way, the two-fermion state (\ref{psift}) is expressed as the 6-level state (\ref{estadof}), with degenerate spectrum.

Introducing Eqs. (\ref{ces}) into (\ref{concurrencef}), the (squared) fermionic concurrence writes in terms of the distribution $\{p_n\}$ and the phases $\{\theta_n\}$ as
\begin{eqnarray}\label{concfr}
 C^2_f(\ket{\psi})&=&8^2|w_{12}w_{34}-w_{13}w_{24}+w_{14}w_{23}|^2\\
 &=&4[p_1p_6+p_2p_5+p_3p_4-2\sqrt{p_1p_6p_2p_5}\cos\alpha+\nonumber\\
 &&+2\sqrt{p_1p_6p_3p_4}\cos\beta-2\sqrt{p_2p_5p_3p_4}\cos(\beta-\alpha)],\nonumber
\end{eqnarray}
with
\begin{subequations}
\begin{eqnarray}\label{angles}
\alpha&=&\theta_{1}+\theta_{6}-\theta_{2}-\theta_{5},\\
\beta&=&\theta_{1}+\theta_{6}-\theta_{3}-\theta_{4}.
\end{eqnarray}
\end{subequations}
\section{Orthogonality condition}\label{ortogonal}

We now assume that the system attains an orthogonal state at $t=\tau$, so the \emph{orthogonality condition}
\beq\label{traslape}
\langle \psi(0)|\psi(\tau)\rangle=\sum_{n}p_ne^{-iE_n\tau/\hbar}=0
\eeq
holds. This is equivalent to 
\beq\label{traslape0}
p_1+p_2\,\chi+(p_3+p_4)\,\chi^2+p_5\,\chi^3+p_6\,\chi^4=0,
\eeq
where 
\beq\label{phisol}
\chi=e^{-i\phi},\quad \phi=\epsilon\tau/\hbar. 
\eeq
Multiplication of (\ref{traslape0}) by $\chi^{*2}$ gives, after separation of the resulting equation into its real and imaginary parts, the couple of conditions 
\begin{subequations}\label{system}
\begin{eqnarray}
\!\!0\!\!&=&\!\!(p_3+p_4)+(p_2+p_5)\cos\phi+(p_1+p_6)\cos2\phi,\\
\!\!0\!\!&=&\!\!(p_2-p_5)\sin\phi+(p_1-p_6)\sin2\phi.
\end{eqnarray}
\end{subequations}
For convenience we now introduce the variables
\begin{subequations}\label{newvar}
\begin{align}
x&=p_1+ p_6,\quad y=p_1-p_6,\\
u&=p_2+ p_5,\quad v=p_2-p_5,\\
w&=p_3+p_4,\quad z=p_3-p_4,
\end{align}
\end{subequations}
in terms of which the normalization constraint reads
\beq \label{norm2}
x+u+w=\sum^6_{n=1}p_n=1,
\eeq
and (\ref{system}) rewrites as
\begin{subequations}\label{system2}
\begin{eqnarray}
\!\!0\!\!&=&\!\!2x\cos^2\phi+u\cos\phi+(1-2x-u),\label{real1}\\
\!\!0\!\!&=&\!\!\big(v+2y\cos\phi\big)\sin\phi\label{im1}.
\end{eqnarray}
\end{subequations}
The couple of equations (\ref{system2}) is equivalent to the orthogonality condition (\ref{traslape}). Clearly it imposes simultaneous restrictions on both the {\it orthogonality time} $\tau=\epsilon\phi/\hbar$ and the variables (\ref{newvar}), which in turn impose conditions on the distribution $\{p_n\}$. Below we analyze in detail such constraints.


\section{Quantum speed limit}\label{secqsl}


The orthogonality time $\tau$ is tightly bounded by the so-called quantum speed limit $\tau_{\textrm{qsl}}$,  \cite{LevitinPRL2009}
\beq\label{Tmin}
\tau\geq\tau_{\textrm{qsl}}=\max\Big\{\frac{\pi\hbar}{2(\langle\hat H\rangle-E_{\min})},\frac{\pi\hbar}{2\sigma_H}\Big\}.
\eeq
Here $\langle\hat H\rangle$ is the mean energy of the system in the state $\ket{\psi}$, $\sigma_H$ its energy dispersion, $\sigma_H=\sqrt{\langle\hat H^2\rangle-\langle\hat H\rangle^2}$, and $E_{\min}$ the lowest energy level, so $\langle\hat H\rangle-E_{\min}$ stands for the mean energy relative to the ground state. The fundamental limit $\tau_{\textrm{qsl}}$ is a unified one that comprises the Mandelstam-Tamm (MT) \cite{Mandelstam1991}, and the Margolus-Levitin (ML) \cite{MargolusLevitinPhysD1998} bounds, respectively given by
\beq\label{MM}
\tau_{\tiny{\rm{MT}}}=\frac{\pi\hbar}{2\sigma_H},\quad \tau_{\tiny{\rm{ML}}}=\frac{\pi\hbar}{2(\langle\hat H\rangle-E_{\min})}.
\eeq
Though it is customary to fix $E_{\min}=0$, so the ML bound is usually expressed as $=\pi\hbar/2\langle\hat H\rangle$, a tighter bound is obtained by taking $E_{\min}$ as the minimal energy actually attainable by the system in the state under consideration. Therefore $E_{\min}=\min{\{E_n|\,p_n\neq0\}}$, and consequently for the equally-spaced case considered here we have $E_{\min}=\epsilon m$, with
\beq\label{emes}
m= \begin{cases}
  3  & p_1\neq0,\\
  4 & p_1=0,\, p_2\neq0,\\
  5&p_1=p_2=0,\,p_3+p_4\neq0,\\
  6&p_1=p_2=p_3=p_4=0,\,p_5\neq 0,\\
  7&p_1=p_2=p_3=p_4=p_5=0,\,p_6=1,
 \end{cases}
\eeq
and
\begin{subequations}\label{means}
\begin{eqnarray}
\langle\hat H\rangle-E_{\min}&=&\epsilon[(5-m)-(2y+v)],\\
\sigma_{H}&=&\epsilon\sqrt{4x+u-(2y+v)^2}.\label{sigma}
\end{eqnarray}
\end{subequations}
Notice from these and Eq. (\ref{Tmin}) that the quantum speed limit depends explicitly on the probabilities $\{p_n\}$ via only {\it two} variables, $4x+u$ and $2y+v$.

\begin{figure}[!]
	\vspace{0.4cm}
	\begin{center}
		\includegraphics[width=0.8\columnwidth]{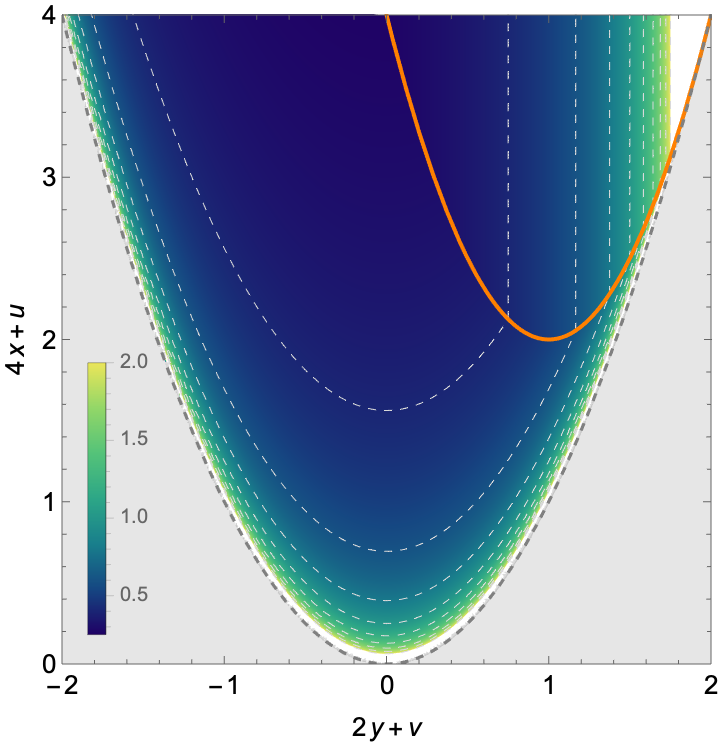}
		\caption[]{\label{alpha} Quantum speed limit $\tau_{\rm{qsl}}$ in the space $(2y+v,4x+u)$ for $p_1\neq 0$ (color scale indicates $\tau_{\rm{qsl}}$ in units of $\pi\hbar/\epsilon$, ranging from $1/4$ to $2$. Higher values are found in the white areas). The orange curve contains distributions for which $\tau_{\rm{qsl}}=\tau_{\rm{ML}}=\tau_{\rm{MT}}=1$, while the region above [below] the curve corresponds to $\tau_{\rm{qsl}}=\tau_{\rm{ML}}$ [$\tau_{\rm{qsl}}=\tau_{\rm{MT}}$]. The dashed gray parabola indicates an infinite orthogonality time.}
	\end{center}
\end{figure}

Since $0\leq\langle\hat H\rangle-E_{\min}\leq E_{\max}-E_{\min}$, with $E_{\max}=\max\{E_n|\,p_n\neq 0\}=\epsilon M$ and
\beq
M= \begin{cases}
  7  & p_6\neq0,\\
  6 & p_6=0,\, p_5\neq0,\\
  5&p_6=p_5=0,\,p_3+p_4\neq0,\\
  4&p_6=p_5=p_4=p_3=0,\,p_2\neq 0,\\
  3&p_6=p_5=p_4=p_3=p_2=0,\,p_1\neq 0,
 \end{cases}
\eeq
it holds that $0\leq\langle\hat H\rangle-E_{\min}\leq\epsilon(M-m)$, whence
\beq \label{cotaAa}
-2\leq 5-M\leq 2y+v\leq 5-m\leq 2.
\eeq 
In its turn, the normalization condition (\ref{norm2}) restricts the domain of $4x+u$ according to 
\beq \label{cotaAb}
0\leq 4x+u\leq4.
\eeq 
With the above expressions we can analyze the ratio $\tau_{\rm{ML}}/\tau_{\rm{MT}}$ in the parameter space $(2y+v,4x+u)$, and map the regions for which the quantum speed limit is given by the MT or the ML bound. Figure \ref{alpha} shows such regions for $m=3$, so the states considered include at least a component along the energy eigenstate with the lowest eigenvalue, and $E_{\min}=3\epsilon$. (Notice that in passing from the variables $4x+u, 2y+v$ to the probabilities $\{p_n\}$ additional constrains must be introduced, namely that $0\leq p_n\leq 1, \sum_np_n=1$, so not all the points in Fig. \ref{alpha}) are associated with a legitimate distribution, yet of course any legitimate distribution is represented in it).

The orange parabola corresponds to $\tau_{\rm{ML}}/\tau_{\rm{MT}}=1$, and comprises all the distributions $\{p_n|\,p_1\neq 0\}$ for which the energy dispersion equals the relative mean energy. It contains, in particular, the distributions $\{p_n|\,p_1=p_m=1/2, m=2,3,4,5,6\}$, corresponding to the two-level (effective qubit) states
\beq\label{cubit}
|\psi^{(1,m)}_{\textrm{q}}\rangle=\frac{1}{\sqrt 2}(\ket{\boldsymbol{1}}+e^{i\varphi}\ket{\boldsymbol{m}}),
\eeq
which, in addition, saturate the quantum speed limit, hence satisfy $\tau=\tau_{\rm{qsl}}=\tau_{\rm{MT}}=\tau_{\rm{ML}}$ \cite{LevitinPRL2009}. We will come back to these two-level states in subsection \ref{qubit}.

The region above the orange parabola corresponds to $\tau_{\rm{ML}}/\tau_{\rm{MT}}>1$, thus contains distributions characterizing states with $\tau_{\textrm{qsl}}=\tau_{\rm{ML}}$. Points below the orange curve correspond to $\tau_{\rm{ML}}/\tau_{\rm{MT}}<1$, so the associate states have $\tau_{\textrm{qsl}}=\tau_{\rm{MT}}$. The density plots inside each zone indicate the value of $\tau_{\textrm{qsl}}$ for each point in it, in units of $\pi\hbar/\epsilon$. Lighter regions represent higher values of the quantum speed limit, here plotted up to a value of 2; points with $\tau_{\textrm{qsl}}\geq 2\pi\hbar/\epsilon$ fill the white areas. In particular, stationary states, with $\sigma_H=0$ and points lying on the gray dashed parabola, do have an infinite value of $\tau_{\textrm{qsl}}$ hence do not attain an orthogonal state. Notice that the point $(2,4)$ in the plane, representing the distribution $\{p_n=\delta_{1n}\}$, locates at the intersection of the dashed-grey and the orange parabolas, since in this case $\tau_{\textrm{qsl}}=\tau_{\textrm{ML}}=\tau_{\textrm{MT}}\rightarrow \infty$. 

As seen in Fig. \ref{alpha} the states with lower values of $\tau_{\textrm{qsl}}$ are those with $2y+v=0$ and high values of $4x+u$. Particularly, the state with the lowest quantum speed limit, namely $\tau_{\textrm{qsl}}=\pi\hbar/4\epsilon$, is represented by the point $(0,4)$ and corresponds to $p_1=p_6=1/2$, i.e, to the qubit state (\ref{cubit}) with $m=6$. Since (as stated above) in this case the orthogonality time coincides with $\tau_{\textrm{qsl}}$, such state attains an orthogonal one faster than any other initial configuration of the system.

\section{Orthogonality time and entanglement for specific distributions}\label{examples}
Before analyzing in detail the orthogonality condition (\ref{system2}), and to get some first insight into the relation (if any) between the speed of evolution and the fermion-fermion entanglement, we will focus on particular distributions $\{p_n\}$, determine the corresponding orthogonality times and analyze the fermionic concurrence of the associate states. 
\subsection{Equiprobable-state distribution}
When all the six energy eigenstates in (\ref{estadof}) are equally probable, $p_n=1/6$ for all $n$ and
\beq
x=u=w=1/3,\quad
y=v=z=0.
\eeq
Since $2y+v$ vanishes, we can immediately conclude from Fig. \ref{alpha} that in this case the quantum speed limit is given by the MT bound. This can be verified resorting to Eqs. (\ref{means}), from which we obtain  
\beq\label{qsl1}
\tau_{\textrm{qsl}}=\tau_{\rm{MT}}=\frac{\pi\hbar}{2\epsilon}\sqrt{\frac{3}{5}}.
\eeq

The orthogonality condition (\ref{system2}) has the solutions (recall that $\phi=\epsilon\tau/\hbar$)
\begin{subequations}\label{sol1}
\begin{eqnarray}
\tau_j&=&\sqrt{\frac{5}{3}}\,(2j+1)\,\tau_{\textrm{qsl}},\quad j=0,1\dots,\\
\tau_k&=&\frac{4}{3}\sqrt{\frac{5}{3}}\,(1+3k)\,\tau_{\textrm{qsl}},\quad k=0,1\dots,\\
\tau_l&=&\frac{4}{3}\sqrt{\frac{5}{3}}\,(2+3l)\,\tau_{\textrm{qsl}},\quad l=0,1\dots,
\end{eqnarray}
\end{subequations}
being
\beq\label{taumin1/6}
\tau_{j=0}=\frac{\pi\hbar}{2\epsilon}
\eeq 
the first time at which an orthogonal state is reached.
\begin{figure}[h!]
	\vspace{0.5cm}
	\begin{center}
		\includegraphics[width=0.9\columnwidth]{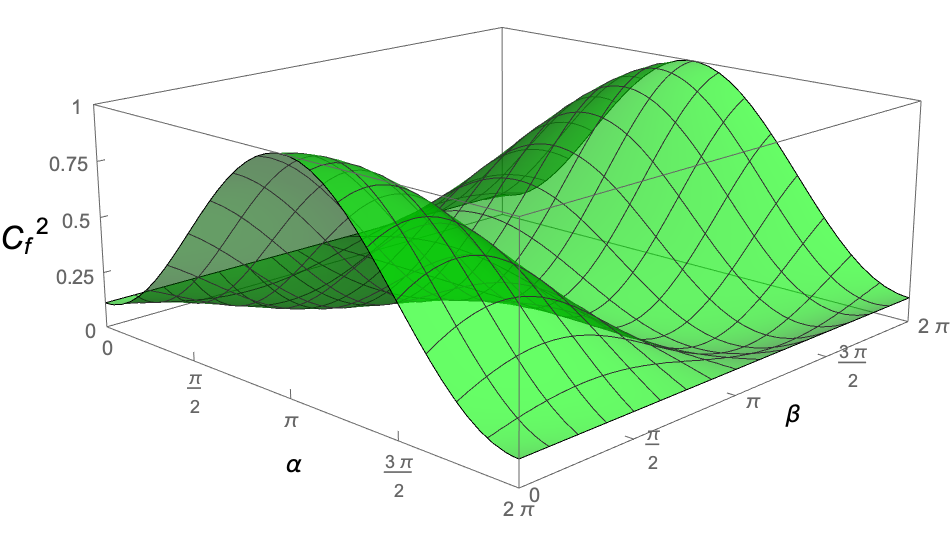}
		\caption[]{\label{Cstate} $C_f^2$ as a function of $\alpha$ and $\beta$ for $p_n=1/6$. The minimum value ($0$) is found at $\alpha=5\pi/3$, $\beta=4\pi/3$, and the maximum ($1$) at $\alpha=\pi$, $\beta=0$.}
	\end{center}
\end{figure}

As for the entanglement of the state (\ref{estadof}) with $p_n=1/6$, Eq. (\ref{concfr}) gives
\beq\label{cf1}
C^2_f=\frac{1}{9}\big\{3-2[\cos \alpha-\cos \beta+\cos(\beta-\alpha)]\big\}. 
\eeq
Figure \ref{Cstate} shows the (squared) fermionic concurrence as a function of $\alpha$ and $\beta$. It can acquire values in the entire interval $[0,1]$, so by an appropriate selection of the phases $\{\theta_n\}$, states with the same energetic resources can be prepared that have any possible amount of entanglement and attain an orthogonal state at times given by (\ref{sol1}).
\subsection{Equiprobable-energy distribution}
Due to the degeneracy of the states $\ket{\boldsymbol{3}}$ and $\ket{\boldsymbol{4}}$, the state-probability distribution $\{p_n\}$ differs from the energy-probability distribution $\{\mathcal{P}_E\}$, the latter given by
\begin{subequations}
\begin{eqnarray}
\mathcal{P}_{E_n}&=&p_n\quad \textrm{for }\quad n=1,2,5,6,\\
\mathcal{P}_{E_3=E_4}&=&p_3+p_4.
\end{eqnarray}
\end{subequations}
When all the energies are equally probable we have $\mathcal{P}_{E}=1/5$ for all $E$. In such case we get
\beq\label{eqE}
x=u=2w=2/5,\quad
y=v=0,
\eeq
and again the fact that $2y+v$ vanishes implies (from Fig. \ref{alpha}) that $\tau_{\textrm{qsl}}=\tau_{\rm{MT}}$. Direct calculation gives $\sigma_H=\sqrt2\epsilon$, and therefore 
\beq
\tau_{\textrm{qsl}}=\tau_{\rm{MT}}=\frac{\pi\hbar}{2\epsilon}\frac{1}{\sqrt2}.
\eeq

Substituting (\ref{eqE}) into (\ref{system2})) we find the following orthogonality times: 
\begin{subequations}\label{taus1/5}
\begin{eqnarray}
\tau^{+}_{l_\pm}\!&=&\!\!\frac{2\sqrt 2}{\pi}\!\Big[\!\!\pm \arccos\Big(\frac{-1+\sqrt5}{4}\Big)\!+\!2\pi l_\pm\Big]\tau_{\textrm{qsl}},\\
\tau^{-}_{m_\pm}\!&=&\!\!\frac{2\sqrt 2}{\pi}\!\Big[\!\!\pm \arccos\Big(\frac{-1-\sqrt5}{4}\Big)\!+\!2\pi m_\pm\Big]\tau_{\textrm{qsl}},
\end{eqnarray}
\end{subequations}
with $l_+,m_+=0,1,2,\dots$, and $l_-,m_-=1,2,\dots$. 
The first time an orthogonal state is reached corresponds to $\tau^+_{l_+=0}$, and is given by
\beq
\tau^+_{l_+=0}=\frac{\hbar}{\epsilon}\arccos\Big(\frac{-1+\sqrt5}{4}\Big)=\frac{2\pi\hbar}{5\epsilon},
\eeq 
as expected when considering an equally-weighted superposition of five non-degenerate, equally-spaced eigenstates \cite{ValdesJPA2020}. Comparison of the last equation with Eq. (\ref{taumin1/6}) gives  $\tau_{j=0}>\tau^+_{l_+=0}$, so the equally-weighted superposition of non-degenerate states ($\mathcal{P}_E=1/5$) reaches an orthogonal state faster than the equally-weighted superposition of (degenerate) energy eigenstates ($p_n=1/6$). This occurs in absolute terms, i.e. by comparing $\tau_{j=0}$ and $\tau^+_{l_+=0}$, and also in relative terms, i.e., by comparison of these times as measured with respect to the corresponding quantum speed limit ($\tau_{j=0}/\tau_{\textrm{qsl}}$ versus $\tau^+_{l_+=0}/\tau_{\textrm{qsl}}$). 


With $p_3+p_4=1/5=p_n$ with $n=1,2,5,6$, Eq. (\ref{concfr}) for $C^2_f$ gives
\begin{align}
C^2_f&=\frac{8}{5^2}(1-\cos\alpha)+4p_3\Big(\frac{1}{5}-p_3\Big)+\\
+&\frac{8}{5}\sqrt{p_3\Big(\frac{1}{5}-p_3\Big)}[\cos\beta-\cos(\beta-\alpha)],
\end{align}
with $p_3\in[0,1/5]$.
\begin{figure}[h!]
	\vspace{0.5cm}
	\begin{center}
		\includegraphics[width=0.9\columnwidth]{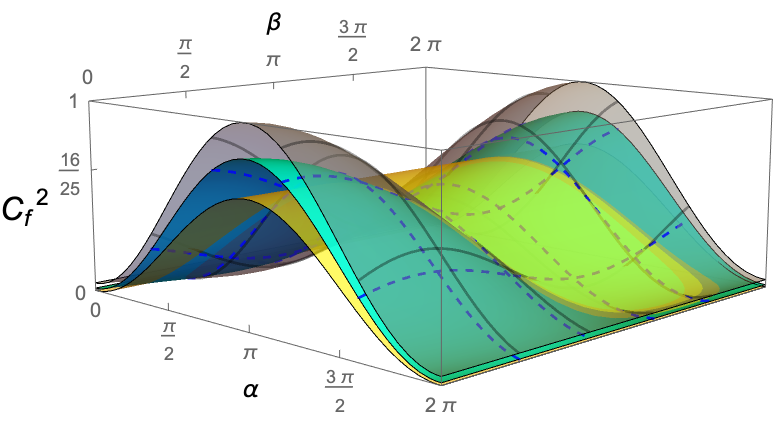}
		\caption[]{\label{concequs} $C_f^2$ as a function of $\alpha$ and $\beta$ for $\mathcal{P}_E=1/5$, and different values of $p_3$: $p_3=0$ (yellow), $p_3=1/60$ (cyan), $p_3=1/10$ (gray).}
	\end{center}
\end{figure}
Figure \ref{concequs} shows the surfaces $C_f^2(\alpha,\beta)$ for different values of $p_3$. For $p_3=0$ (yellow surface) we have $0\leq C_f \leq 4/5$, and as $p_3\rightarrow 1/10$ the concurrence can attain higher values, until for $p_3=1/10$ (gray surface) we have $1/5\leq C_f \leq 1$. As $p_3$ increases from $1/10$ to $1/5$ the inverse behavior occurs (the gray surface transforms into the yellow one). 

We therefore see that in the equiprobable-energy case, initial states with different populations in $\ket{\boldsymbol{3}}$ (or $\ket{\boldsymbol{4}}$) can be prepared that have common orthogonality times (\ref{taus1/5}) and different amounts of entanglement. In particular ---and again with an appropriate choice of the relative phases--- separable states can evolve towards an orthogonal one (for $p_3=0$) as fast as maximally entangled states do (for $p_3=1/10$). Further, by varying $\alpha$ and $\beta$ the fermionic concurrence can be fixed to any value in $[1/5,4/5]$ irrespective of $p_3$. 

\subsection{Effective qubit states}\label{qubit}
This case corresponds to distributions in which only two probabilities, say $p_i$ and $p_j=1-p_i$, are nonzero.
Without loss of generality we assume that $E_i< E_j$, so $E_{\min}=E_i$. We exclude from the analysis the degenerate case with $E_i=E_j$, since in such (stationary) situation the state will never evolve towards an orthogonal one. 

In order to maintain the generality in the indices $i$ and $j$, we resort to the orthogonality condition in the form (\ref{traslape}) instead of (\ref{system2}). Direct calculation shows that in the present two-level case Eq. (\ref{traslape}) can only be satisfied for $p_i=1/2=p_j$, with orthogonality times given by
\beq\label{tauqubit}
\tau_l=\frac{\pi\hbar}{(E_j-E_i)}(2l-1), \quad l=1,2,\dots.
\eeq
With $p_i=p_j=1/2$ clearly $\langle\hat H\rangle-E_{\min}=\langle\hat H\rangle-E_i=\sigma_{H}$, whence 
\beq
\tau_{\rm{qsl}}=\tau_{\rm{MT}}=\tau_{\rm{ML}}=\frac{\pi\hbar}{(E_j-E_i)}=\frac{\pi\hbar}{\epsilon(M-m)},
\eeq
 and (\ref{tauqubit}) rewrites as
\beq
\tau_l=(2l-1)\,\tau_{\rm{qsl}},
\eeq
in line with the well-known results for the effective qubit case \cite{LevitinPRL2009,ValdesJPA2020} which state, in particular, that a first orthogonal state is reached at $\tau_1=\tau_{\rm{qsl}}$, i.e., as fast as permitted by the fundamental quantum speed limit (a property \emph{only} ascribable to qubit systems \cite{LevitinPRL2009}). Further, since $\tau_{1}$ depends on the energy difference $E_j-E_i$, the fastest qubit (in absolute terms) is that corresponding to the state with maximal dispersion, i.e, with $i=1$ and $j=6$, which is precisely the state represented in the point $(0,4)$ in Fig. \ref{alpha}.

As for the entanglement of the qubits that attain orthogonality, it follows from Eq. (\ref{concfr}) that the only ones with non-vanishing $C_f$ are maximally entangled ($C_f=1$), and are those corresponding to $p_1=p_6=1/2$, or to $p_2=p_5=1/2$. Consequently, the states that saturate the quantum speed limit have \emph{extreme} values of entanglement: they are either disentangled or maximally entangled.  

\begin{figure}[h!]
	\vspace{0.5cm}
	\begin{center}
		\includegraphics[width=0.8\columnwidth]{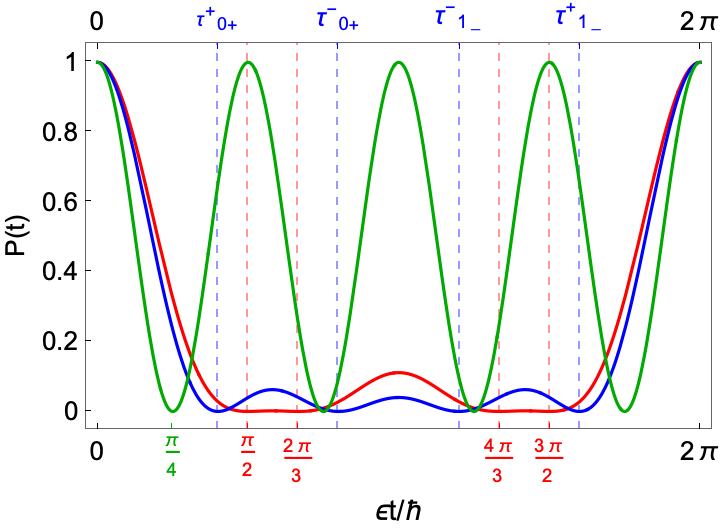}
		\caption[]{\label{distrib} Survival probability $P(t)=|\langle\psi(0)|\psi(t)\rangle|^2$ for different distributions: $p_n=1/6$ (red), $\mathcal{P}_E=1/5$ (blue), and $p_1=p_6=1/2$ (green). The first four orthogonality times are shown.}
	\end{center}
\end{figure}

Before closing this section, we show in Figure \ref{distrib} the evolution of the survival probability $P(t)=|\langle\psi(0)|
\psi(t\rangle|^2$ for the cases considered above: equiprobable-state distribution (red curve), equiprobable-energy 
distribution (blue curve) and two-level state with maximal dispersion (green curve). In line with the previous discussion, the 
qubit reaches an orthogonal state first, at $\tau_1=\pi\hbar/4\epsilon$, followed by the equiprobable-energy superposition, 
and finally by the equiprobable-state superposition. These later states spend considerable
time sufficiently `far away' from the initial state before returning to the initial situation at $\epsilon t/\hbar=2\pi$. Indeed, though not clearly appreciated in the red curve of Fig. 
\ref{distrib}, for $\epsilon t/\hbar$ in the intervals $[\pi/2,2\pi/3]$ and $[4\pi/3,3\pi/2]$, the survival probability lies between $0$ (at the intervals' extreme points) and $\approx0.0017$ (at the center of each interval), 
so during these time windows the system is found in states that are basically distinguishable from the initial one.

In the equiprobable-energy case (blue curve), $P(t)$ results independent of the free parameter $p_3$ (or $p_4$), so the dynamics of the survival probability is the same as that of a 5-level system (vanishing $p_3$ or $p_4$, and $p_i=1/5$ for the rest of the indices $i$) with an equally-spaced spectrum \cite{ValdesJPA2020}.  

\section{Solution regions}\label{regions}

We now focus on the orthogonality condition in the form (\ref{system2}) and identify the restrictions it imposes on the parameter spaces $(x,u)$ and $(y,v)$, which in turn impose further constraints on the distributions $\{p_n\}$, and therefore on the initial states that are allowed to evolve towards an orthogonal one.

First we consider Eq. (\ref{real1}). Its solutions, taking into account the normalization condition (\ref{norm2}), read
\beq\label{chicharronero}
\cos\phi=\begin{cases}
  \frac{-u\pm\sqrt{(4x+u)^2-8x}}{4x}  & x\neq0,\\
  \frac{u-1}{u} & x=0.
 \end{cases}
 \eeq
(Notice that (\ref{norm2}) implies that in the case $x=0$, $u$ is necessarily nonzero, otherwise $x=0=u$ leads to $w=1$, which corresponds to a stationary state that does not evolve towards an orthogonal state.) Clearly, the right-hand sides of (\ref{chicharronero}) constitute indeed a solution provided they acquire values in the interval $[-1,1]$. The upper bound is always satisfied by the expressions in \ref{chicharronero}, so for $x\neq 0$ the imposed restrictions read 
\begin{subequations}\label{condchich}
\begin{eqnarray}
8x\leq(4x+u)^2,\\
u-4x\leq \pm\sqrt{(4x+u)^2-8x},\label{raiz}
\end{eqnarray}
whereas for $x=0$ we have
\end{subequations}
\beq
\frac{1}{2}\leq u.
\eeq 
\begin{figure}[h!]
	\vspace{0.35cm}
		\includegraphics[width=0.5\columnwidth]{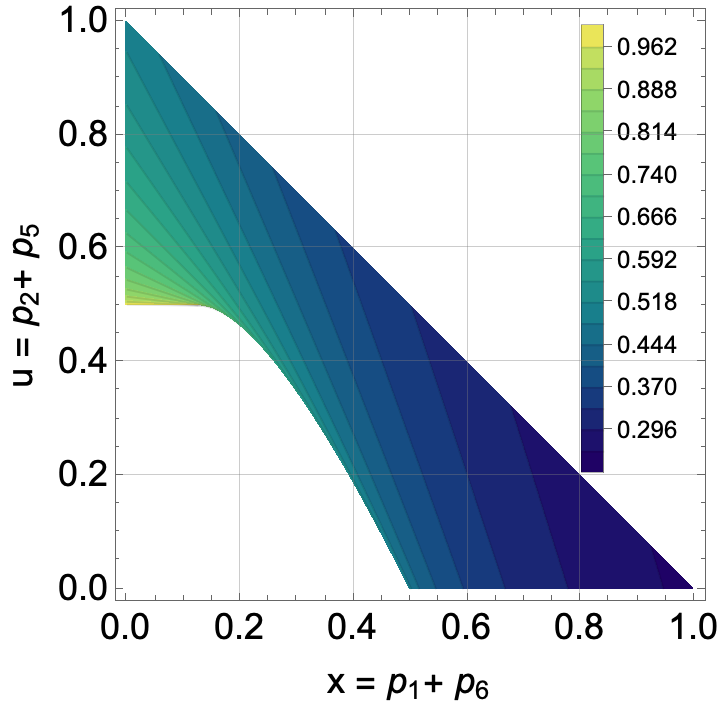}\includegraphics[width=0.5\columnwidth]{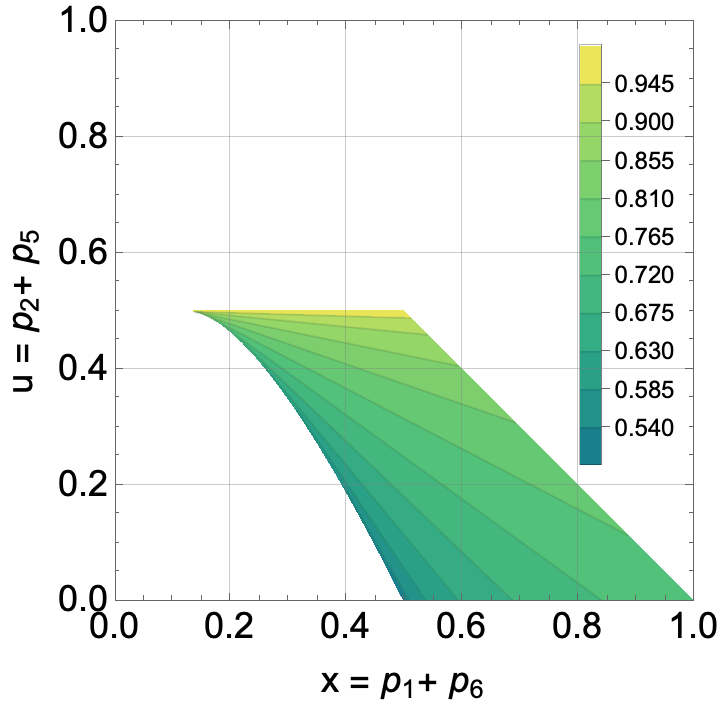}		\caption[]{\label{region xu} Regions in the space $(x,u)$ for which solutions of (\ref{real1}) exist, and corresponding solutions $\phi_1\in(0,\pi]$ in color scale in fractions of $\pi$ (ranging from $1/4$ to $1$). Left panel: Solutions (\ref{chicharronero}) with the plus (upper) sign in the fist line. Right panel: Solutions for $x\neq 0$ with the minus (lower) sign.}
\end{figure}
Further, in all cases it must hold that $0<x+u\leq 1$. 

The regions in the space $(x,u)$ satisfying the above conditions are shown in Fig. \ref{region xu}, and the first solution for $\phi$ ---denoted as $\phi_1$, with values in $(0,\pi]$--- at each point is represented by the contour plot over such regions, scaled with $\pi$. The left panel of Fig. \ref{region xu} corresponds to the solutions in Eq. (\ref{chicharronero}) considering the plus (upper) sign in the first line, and $\phi_1$ is found to lie in the interval $\pi[1/4,1]$. The right panel corresponds to the solutions with $x\neq0$ and the minus (lower) sign in front of the square root, and $\phi_1$ lies in $\pi[1/2,1]$. It thus follows that the first solution of (\ref{real1}) is bounded by
\beq\label{bound}
\frac{\pi}{4}\leq\phi_1\leq\pi.
\eeq
The first inequality is saturated only for $x=1,u=0$, or equivalently for $p_1=p_6=1/2$, whereas the second inequality is saturated only along the line $u=1/2$ and $x\in[0,1/2]$ (for $0\leq x\leq 1/8$ the corresponding points pertain to left panel in Fig. \ref{region xu}, while for $1/8< x\leq 1/2$ they pertain to the right panel). 

The points for which $\phi_1=\pi$ constitute real roots of Eq. (\ref{traslape0}). Indeed, all the real roots of (\ref{traslape0}) are of the form $\phi_l=l\pi$ with $l$ integer; in this case Eq. (\ref{im1}) is trivially satisfied, the orthogonality condition reduces to Eq. (\ref{real1}) only, and consequently $\{\tau_l=(\hbar/\epsilon)\phi_l\}$ defines the set of orthogonality times associated to the real roots of Eq. (\ref{traslape0}). The case $l$ even is ruled out by virtue of Eqs. (\ref{traslape0}) and (\ref{norm2}), hence 
\begin{subequations}\label{famI}
	\beq\label{realtau}
	\tau_l=\frac{\pi\hbar}{\epsilon}l, \quad l=\textrm{odd},
	\eeq
provided (as follows from Eq. (\ref{real1}) and the normalization constraint)
\beq\label{erresA}
u=\frac{1}{2}=x+w.
\eeq
\end{subequations}
Equation (\ref{famI}) means that the yellow segment in Fig. \ref{region xu} ($u=1/2,0\leq x\leq 1/2$) contains {\it all} the real roots of the orthogonality condition, though in Fig. \ref{region xu} only the solutions for $l=1$ are depicted. Further, as we have seen points along the yellow line represent the states that take the longest time to reach an orthogonal one for the first time. 

Since the orthogonality time $\tau=\hbar\phi/\epsilon$ solves \emph{simultaneously} both Eq. (\ref{real1}) and (\ref{im1}) for the same set of parameters, the solutions $\phi_1$ in Fig. \ref{region xu} cannot be all identified with orthogonality times until the solutions of Eq. (\ref{im1}) are also analyzed. While Eq. (\ref{real1}) forbids certain regions of the space $(x,u)$, Eq. (\ref{im1}) imposes restrictions on the parameter space $(y,v)$, and thereby introduces stronger constraints on the probabilities $p_1,p_2,p_5,p_6$. Since the structure of Eq. (\ref{im1}) naturally splits its solutions into the families
\begin{itemize}
\item [I)]solutions for which $\sin\phi=0$,
\item [II)] solutions for which $v+2y\cos\phi=0$,
\end{itemize}
such extra constraints will arise only in the family II. Solutions pertaining to I clearly impose no additional restriction on $y$ or $v$, and constitute the real roots of Eq. (\ref{traslape0}), which have just been identified with a line segment in the plane $(x,u)$. Consequently it only remains to analyze the equation
\beq\label{famII}
v+2y\cos\phi=0
\eeq
and the allowed domain for the parameters $y$ and $v$. The latter is determined by noticing that according to (\ref{newvar}), $v,y$ and $v\pm y$ lie in the interval $[-1,1]$. Equation (\ref{famII}) further imposes $-1\leq -v/(2y)\leq1$, provided $y\neq 0$ (the case $y=0$ will be commented separately below). 

\begin{figure}[h!]
	\vspace{0.35cm}
	\begin{center}
		\includegraphics[width=0.7\columnwidth]{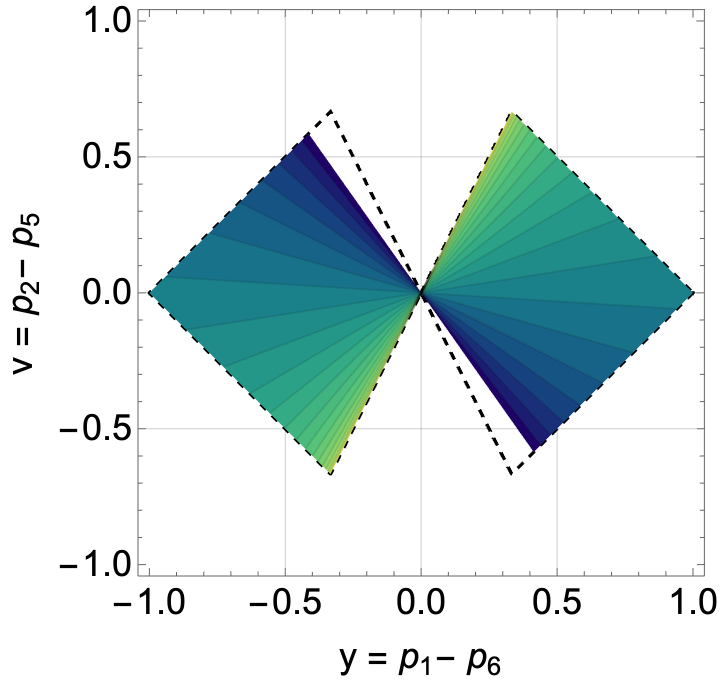}
		\caption[]{\label{region yv} Regions in the space $(y,v)$ for which solutions of (\ref{famII}) exist (delimited by the dashed line), and corresponding first solutions $\phi_1$ in the interval $\pi[1/4,1]$ (color scale in fractions of $\pi$). White regions inside the dashed line are excluded since correspond to $\phi_1<\pi/4$, hence to solutions inconsistent with the orthogonality condition. }
	\end{center}
\end{figure}

The resulting permitted region in the plane $(y,v)$ is delimited by the dashed line in Fig. \ref{region yv}, and the corresponding solutions $\phi_1\in(0,\pi]$ of (\ref{famII}) are shown in color scale in fractions of $\pi$, considering only values within the interval $[1/4,1]$. Values of $\phi_1< \pi/4$, though valid solutions of (\ref{famII}), do not comply with the bound (\ref{bound}) imposed by Eq. (\ref{real1}), and fill the points in the white triangular zones inside the dashed region, excluded as admissible solutions of the orthogonality condition. 

The points with $\phi_1=\pi$ lie along the (yellow) line $v=2y$, which comprises the points at the intersection of the solution families I and II. As for the points for which $\phi_1=\pi/4$, lying along the darker blue edges (with equation $v=-\sqrt{2}y$), the only one that represents a legitimate solution to the orthogonality condition (is solution of \emph{both} equations (\ref{real1}) and (\ref{im1})) is the origin, since as stated below (\ref{bound}), for $\phi_1=\pi/4$ it holds $p_1=p_6=1/2$, which implies $y=0=v$. Notice that the origin of the plane $(y,v)$ represents the only (and trivial) solution to (\ref{famII}) for $y=0$, so at this point the orthogonality times are completely determined from (\ref{chicharronero}) and Fig. \ref{region xu}.

The permitted regions and solution maps shown in Figs. \ref{region xu} and \ref{region yv} offer insight into the structure of the initial states (via the allowed $\{p_n\}$) that reach an orthogonal one, and into the corresponding first orthogonality times. To completely determine such states and times both figures should be analyzed simultaneously; each of them provides information regarding only one of the coupled equations in (\ref{system2}), yet as emphasized before, the actual solutions of the orthogonality condition simultaneously solve (\ref{real1}) and (\ref{im1}). However, inspection of Figs. \ref{region xu} and \ref{region yv} provides qualitative information that allows us to conclude, for example, that faster states are characterized by a low population of the states $\ket{\boldsymbol{2}}$ and $\ket{\boldsymbol{5}}$ (so $u$ and $v$ are small), and a higher (and comparable) population of the extreme states $\ket{\boldsymbol{1}}$ and $\ket{\boldsymbol{6}}$ (meaning $x$ large and $y$ small). In contrast, the slower states have $\ket{\boldsymbol{2}}$ and $\ket{\boldsymbol{5}}$ sufficiently populated, so that there is approximately $50\%$ chance of the system being in either one of these states. 

\section{Entanglement and speed limit}\label{entanglement}
The examples presented in Sec. \ref{examples} show that the dependence of $C_f$ on the relative phases $\alpha$ and $\beta$ may allow for a wide range of values of entanglement for states with the same speed of evolution. A search of the connection between the amount of fermionic entanglement and the relative speed towards orthogonality ---as measured by the \emph{relative orthogonality time} $\tau_1/\tau_{\textrm{qsl}}\geq 1$---, was previously advanced in \cite{BatlePRA2005} for the system here under consideration, but restricted to the solution family II, with $y=0$. No such connection was found. Later on the assumption of an equally-spaced Hamiltonian was relaxed \cite{OliveiraIJQ2008}, and it was shown that a relation between $C_f$ and $\tau_1/\tau_{\textrm{qsl}}$ may indeed emerge for an appropriate energy spectrum. 

\begin{figure}[h]     
    \centering
       \includegraphics[width=0.8\columnwidth,  trim= 130pt 70pt 130pt 70pt]{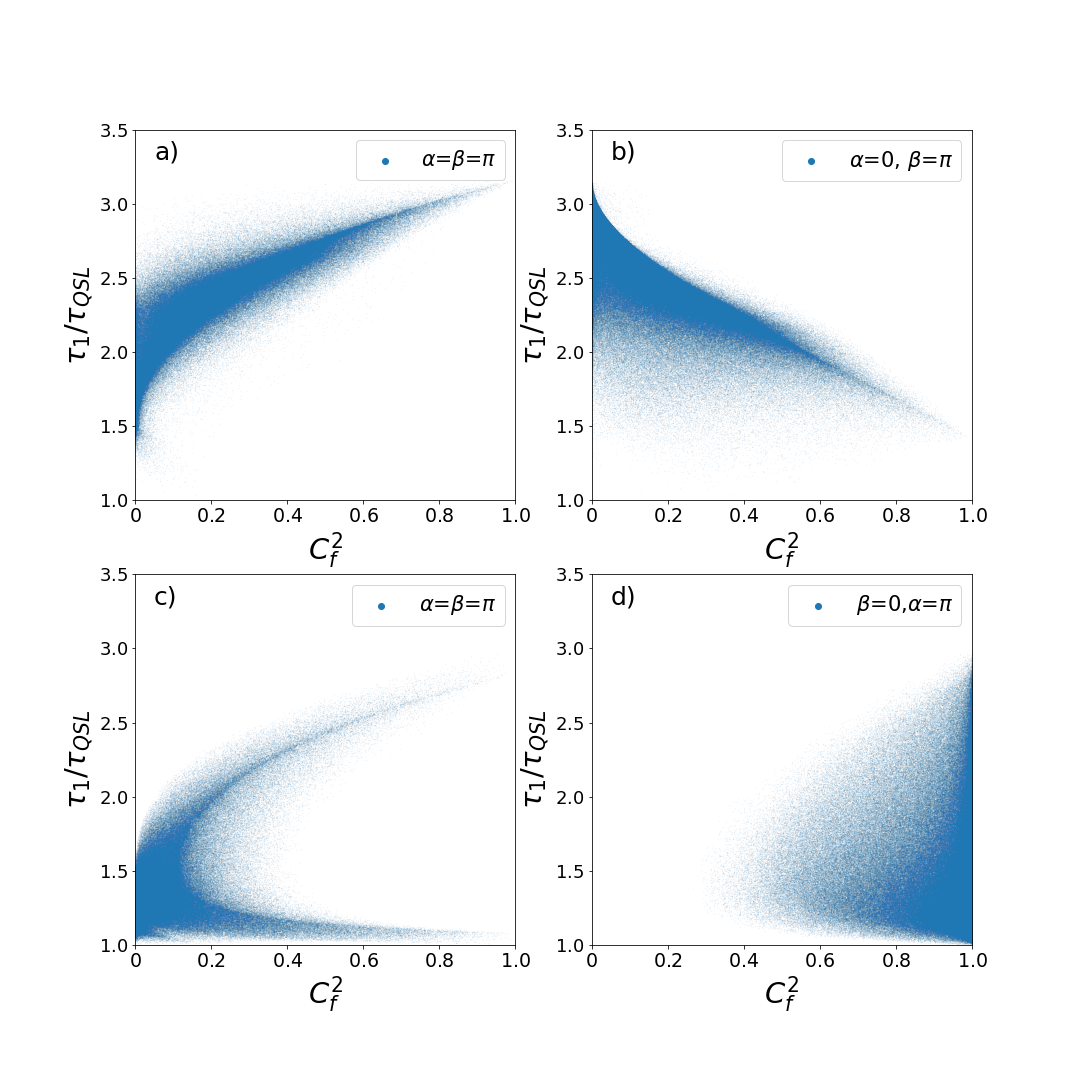}
   \caption{Points in the plane $(C^2_f,\tau_1/\tau_{\textrm{qsl}})$ representing four classes of states that reach an orthogonal one for the first time at $\tau_1$. Each class (panel) is characterized by a distribution $\{p_n\}$ and a pair of phases $(\alpha,\beta)$ according to: a) Solutions in family I, with $\alpha=\beta=\pi$; b) Solutions in family I, with $\alpha=0,\beta=\pi$; c) Solutions in family II for $y\neq 0$, with $\alpha=\beta=\pi$; d) Solutions in family II for $y=0$, with $\alpha=\pi,\beta=0$. In each panel 300,000 points are plotted.}
\label{puntos}
\end{figure}

To explore in more detail the possible relation between $C_f$ and the relative orthogonality time (in the present equally-spaced spectrum case), we search for 300,000 distributions $\{p_n\}$ consistent with the orthogonality condition; for each of them $\tau_1$ is directly obtained from the solutions of (\ref{system2}), $\tau_{\textrm{qsl}}$ follows from Eqs. (\ref{Tmin}) and (\ref{means}), and (\ref{concfr}) gives $C^2_f$ as a function of $\alpha$ and $\beta$ only. By assigning values to these phases we construct the $\tau_1/\tau_{\textrm{qsl}}$ vs $C^2_f$ plots in Fig. \ref{puntos}, where each point represents the set of states characterized by one distribution $\{p_n\}$ and the corresponding pair $(\alpha,\beta)$. Since throughout this section we do not focus on the absolute (first) orthogonality time $\tau_1$ (as we did before), but rather on the ratio $\tau_1/\tau_{\textrm{qsl}}$, a comment is in place here regarding the usage of the terms `faster' or `slower'. When the relevant time is measured by $\tau_1$, a state $A$ is faster than a state $B$ if $\tau^A_1<\tau^B_1$, meaning that in absolute terms $A$ arrives to an orthogonal state earlier than $B$. When the relative time is employed instead, a state $A$ is said to be faster than the state $B$ if $\tau^A_1/\tau^A_{\textrm{qsl}}<\tau^B_1/\tau^B_{\textrm{qsl}}$, meaning that the speed of $A$ \emph{relative to the limit imposed by the system's energetic limitations} is greater than the corresponding speed of $B$. (Notice that in general  $\tau^A_1<\tau^B_1$ does not imply $\tau^A_1/\tau^A_{\textrm{qsl}}<\tau^B_1/\tau^B_{\textrm{qsl}}$.)
In agreement with \cite{BatlePRA2005} we find that by randomly choosing $\alpha$ and $\beta$ no correlation exists between $C_f$ and $\tau_1/\tau_{\textrm{qsl}}$. However, classes of states can be identified ---characterized by a distribution corresponding to either one of the families I and II, and by specific values of the phases--- in which correlations may arise. To illustrate this, Fig. \ref{puntos} shows the points in the plane $(C^2_f,\tau_1/\tau_{\textrm{qsl}})$ for four classes of states which exhibit qualitatively different relations between the relative orthogonality time and the amount of fermion-fermion entanglement.    
\begin{figure}[h]     
    \centering
       \includegraphics[width=0.8\columnwidth, trim= 120pt 70pt 120pt 70pt]{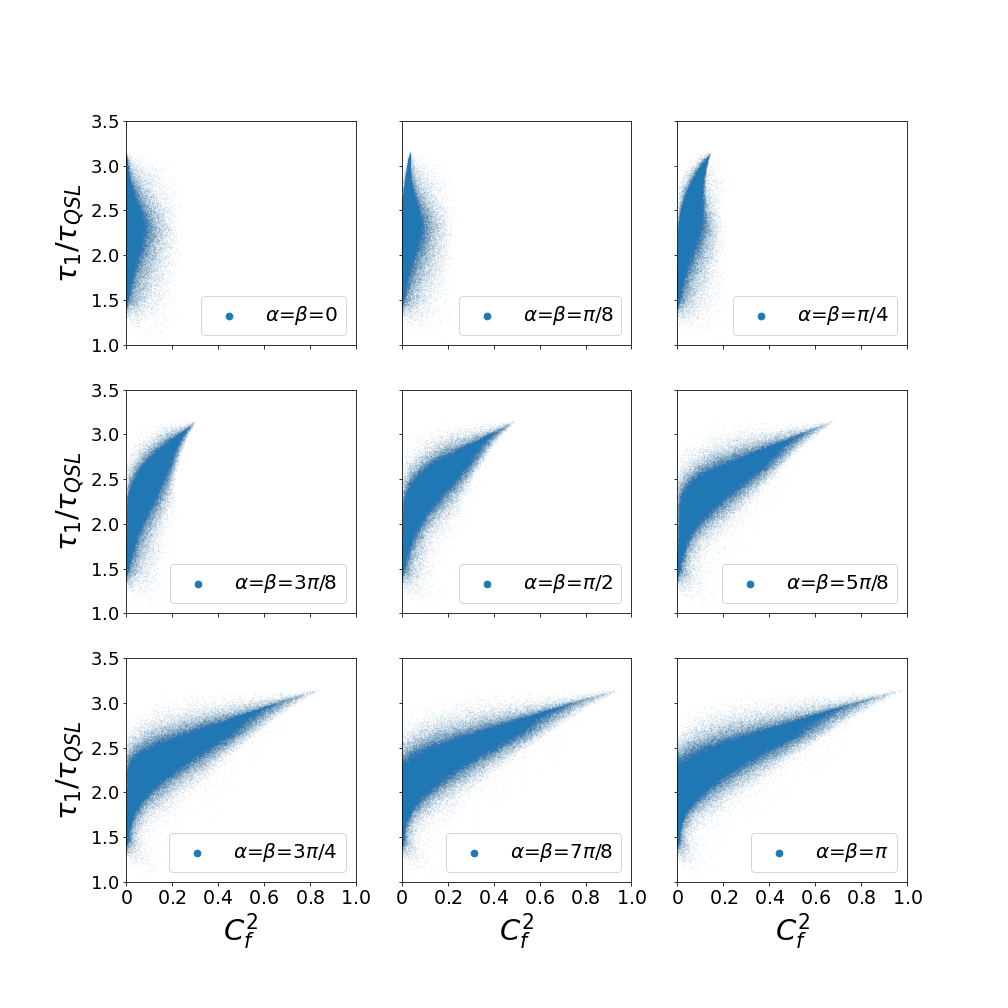}
   \caption{Sequence of the evolution of the points in $(C^2_f,\tau_1/\tau_{\textrm{qsl}})$ for the class of states shown in Fig. \ref{puntos}a) as $\alpha=\beta=\pi$ increases from $0$ (upper left panel) to $\pi$ (bottom right panel). The reversed sequence is obtained as $\alpha=\beta$ varies from $\pi$ to $2\pi$. In each panel 300,000 points are plotted.}
\label{secuencia}
\end{figure}
Figure \ref{puntos}a) corresponds to real solutions of the orthogonality condition (Family I), hence to distributions $\{p_n\}$ represented by yellow points in Fig. \ref{region xu}, and fixed $\alpha=\beta=\pi$. In this case there is a clear tendency of the highly entangled states to be the slower ones, and barely entangled states tend to evolve faster, yet a wide spread in $\tau_1/\tau_{\textrm{qsl}}$ is observed for small $C_f$. Figure \ref{puntos}b) also represents real solutions of (\ref{system2}), but now with fixed $\alpha=0$, $\beta=\pi$, and the previous tendency is reverted: as the entanglement increases, the states tend to speed up towards an orthogonal one. 

States corresponding to solutions in the family II with $y\neq 0$ are depicted in Figure \ref{puntos}c) for $\alpha=\beta=\pi$. An interesting pattern is shown, in which as the fermionic concurrence increases a bifurcation in the corresponding $\tau_1/\tau_{\textrm{qsl}}$ emerges, so highly entangled states transit towards an orthogonal one with an extreme speed, being either very fast or very slow.   
In the last example, shown in Figure \ref{puntos}d), we consider solutions again in the family II but now satisfying $y=0$ (the origin in Fig. \ref{region yv}), with $\alpha=\pi$, $\beta=0$. In contrast with the previous cases, barely entangled states are scarce and highly entangled ones exhibit no relation to the corresponding relative orthogonality time. 

It is clear from Figures \ref{puntos} that the relation between $\tau_1/\tau_{\textrm{qsl}}$ and $C_f$ is not unique, and different clear tendencies may arise determined by the relative phases $\alpha$ and $\beta$. Moreover, such tendencies are continuously outlined as these phases are continuously varied, while preserving a fixed relation between them. This is illustrated in Fig. \ref{secuencia}, in which the points in the plane $(C^2_f,\tau_1/\tau_{\textrm{qsl}})$ for the class of states shown in Fig. \ref{puntos}a) are shown, sequenced as the phases $\alpha=\beta$ increase from $0$ to $\pi$. Therefore, whereas for arbitrary states ---arbitrary $\{p_n\}$ (complying of course with the orthogonality requirements), and arbitrarily chosen $\alpha$ and $\beta$--- a connection between the speed of evolution and the fermion-fermion entanglement is missing, there exist classes of states ---corresponding to $\{p_n\}$ consistent with one of the two families of solutions, and to phases $\alpha$ and $\beta$ related in a suitable way--- in which the connection is disclosed. Different classes of states exhibit different relations between entanglement and quantum speed, a feature that can be exploited with an appropriate preparation of the initial state that guarantees, for example, that the state will be a highly entangled one that attains orthogonality in the largest/shorter allowed time.



\section{Final remarks}\label{conclusions}
We have considered the simplest fermionic system that exhibits the phenomenon of entanglement on top of the exchange correlations, to analyze its speed of evolution towards an orthogonal state, and the relation of this latter with the so-called fermionic entanglement. 

The analysis involving the conditions required to reach an orthogonal state in a finite time relies solely on the energy-decomposition \ref{estadof}, hence applies not only to our fermionic system but also to a wide family of states ---namely those of the form \ref{estadof} with an spectrum given by \ref{spectrum}--- irrespective of the specific nature of the system. Consequently, the results of Sects. \ref{ortogonal}, \ref{secqsl} and \ref{regions} accommodate a wider variety of physical situations that comprises $N$-level systems (with $N\leq 6$ and possible energy degeneracies), besides the two-fermion system under study. For all such cases we have mapped the (distribution-dependent) parameter space $(2y+v,4x+u)$ according to the states' quantum speed limit, thus revealing the regions where $\tau_{\textrm{qsl}}$ increases/decreases as the distribution $\{p_n\}$ is varied (Fig. \ref{alpha}). 

Further, the regions in the spaces $(x,u)$ and $(y,v)$ that admit solutions of each of the equations that conform the orthogonality condition are identified, and the corresponding time $\tau_1$, at which a first orthogonal state is reached, is determined (Figs. \ref{region xu} and \ref{region yv}). Such time is bounded from below by $\pi\hbar/4\epsilon=\tau_1(|\psi^{(1,6)}_{\textrm{q}}\rangle)$, and from above by $\pi\hbar/\epsilon=\tau_1(|\psi^{(i,j)}_{\textrm{q}}\rangle)$, with $(i,j)\in\{(1,2),(2,3),(2,4),(3,5),(4,5),(5,6)\}$. Therefore the qubit states with maximal and minimal (yet non-zero) dispersion are, respectively, the fastest and the slowest states, in an absolute time-scale. 
Qualitative information can be extracted from the color maps in Figs. \ref{region xu} and \ref{region yv}, that highlights the role of the different probabilities $\{p_n\}$, or populations of the energy eigenstates, in leading to a faster or slower state. In this way, our results facilitate to determine which populations should be favored in order to speed up, or speed down, the evolution towards orthogonality.   

As for our analysis of the fermion-fermion entanglement and its relation with the speed of evolution, we found that when the quantum speed limit is saturated, so $\tau_1=\tau_{\textrm{qsl}}$, the entanglement results either maximal ($C_f=1$) or minimal ($C_f=0$), so the entanglement of the states with maximal (relative) speed may acquire only extreme values. For cases in which $\tau_1>\tau_{\textrm{qsl}}$, the examples in Sect. \ref{examples} evince that states with different amounts of entanglement may share the same \emph{absolute} orthogonality times. Further, the numerical analysis in Sect. \ref{entanglement} reveals that states with different fermionic concurrence may share the same \emph{relative} orthogonality times. A direct relation between the entanglement of a given (arbitrary) state and its speed towards orthogonality is therefore missing. Moreover, when a sample of random initial states is considered, no correlation between $C_f$ and $\tau_1/\tau_{\textrm{qsl}}$ is revealed. This confirms the result reported in \cite{BatlePRA2005}; yet, when specific classes of states are taken into consideration (classes for which the relative phases $\theta_n$ in Eq. (\ref{estadof}) are suitably related) a connection between entanglement and the speed of evolution is disclosed. Such connection, however, is not universal and may vary among the different classes, a feature that allows to select specific initial states to induce different correlations between $C_f$ and $\tau_1/\tau_{\textrm{qsl}}$. 

That a connection between entanglement and the speed towards orthogonality may exist not only in systems of distinguishable parties, or in composites of indistinguishable bosons, but also in systems of indistinguishable fermions goes in line with the conclusions reached in \cite{OliveiraIJQ2008}. However, while in \cite{OliveiraIJQ2008} the emergence of such connection was ascribed to a non-equally-spaced spectrum, here we ascribe it to the class of the initial state, without the need to invoke more general energy spectra. These observations indicate that correlations between fermionic entanglement and speed of evolution may exist, yet manifest in more subtle ways.

%

\acknowledgements
The authors acknowledge financial support from DGAPA, UNAM through project PAPIIT IN113720.

\end{document}